\documentclass[fleqn,usenatbib]{mnras}

\usepackage[T1]{fontenc}

\DeclareRobustCommand{\VAN}[3]{#2}
\let\VANthebibliography\thebibliography
\def\thebibliography{\DeclareRobustCommand{\VAN}[3]{##3}\VANthebibliography}

\usepackage{graphicx}	
\usepackage{amsmath}	
\usepackage{amssymb}	
\usepackage{color}
\usepackage{booktabs}
\usepackage{breakcites}
\usepackage[table]{xcolor}

\usepackage{empheq}
\usepackage{newtxtext,newtxmath}
\usepackage[normalem]{ulem}

\mathchardef\mhyphen="2D

\newcommand{\di}{\mathrm{d}}

\newcommand{\pc}{\,{\rm pc}}
\newcommand{\kpc}{\,{\rm kpc}}

\newcommand{\kms}{\,{\rm km\, s^{-1}}}

\newcommand{\Msun}{\, \rm M_\odot}

\newcommand{\rhodisc}{\rho_{\rm disc}}
\newcommand{\rhobara}{\rho_{\rm bar,1}}
\newcommand{\rhobarb}{\rho_{\rm bar,2}}
\newcommand{\rhobarc}{\rho_{\rm bar,3}}

\DeclareMathOperator{\sech}{sech}

\title[The stellar mass distribution of the MW bar]{The stellar mass distribution of the Milky Way's bar: an analytic model}

\author[Sormani et al.]{
Mattia C. Sormani,$^{1}$\thanks{E-mail: mattia.sormani@uni-heidelberg.de}
Ortwin Gerhard,$^{2}$
Matthieu Portail,$^{3}$
Eugene Vasiliev,$^{4}$ 
Jonathan Clarke$^{2}$ 
\\
$^1$ Universit\"{a}t Heidelberg, Zentrum f\"{u}r Astronomie, Institut f\"{u}r theoretische Astrophysik, Albert-Ueberle-Str. 2, 69120 Heidelberg, Germany \\
$^2$ Max-Planck-Institut fur Extraterrestrische Physik, Gie{\ss}enbachstra{\ss}e 1, D-85748 Garching, Germany \\
$^3$ CAPE Analytics, Gottfried-Keller-Straße 35, D-81245 München, Germany \\
$^4$ Institute of Astronomy, University of Cambridge, Madingley Rd, Cambridge, CB3 0HA, UK \\
}

\date{Accepted XXX. Received YYY; in original form ZZZ}
\pubyear{2015}

\begin{document}
\label{firstpage}
\pagerange{\pageref{firstpage}--\pageref{lastpage}}
\maketitle

\begin{abstract}
We present an analytic model of the stellar mass distribution of the Milky Way bar. The model is obtained by fitting a multi-component parametric density distribution to a made-to-measure N-body model of Portail et al., constructed to match a variety of density and kinematics observational data. The analytic model reproduces in detail the 3D density distribution of the N-body bar including the X-shape. The model and the gravitational potential it generates are available as part of the software package {\sc Agama} for galactic dynamics, and can be readily used for orbit integrations, hydrodynamical simulations or other applications.
\end{abstract}



\section{Introduction} \label{sec:introduction}

At the beginning of the 1990s it became established that the Milky Way (MW) is a barred galaxy \citep{Blitz1991,Binney1991,Weiland1994,Stanek1994}. During the following three decades, our knowledge of the dynamical structure of the Galactic bar has vastly increased thanks to near-infrared photometric observations \citep{Binney1997,Launhardt2002,Ness2016}, star counts \citep{Stanek1997,Skrutskie2006,Saito2011,Cao2013,
Wegg2015,
Coleman2020}, line-of-sight velocity \citep{Kunder2012,Nidever2012,Ness2013,Zoccali2014,Bovy2019} and proper motion  \citep{Sanders2019a,Sanders2019b,Clarke2019} data, as well as stellar \citep{Fux1997,Shen2010,Molloy2015,Portail2017a,Portail2017b} and gas dynamical \citep{Fux1999,Bissantz2003,Sormani2015a,Li2016,Li2022} modelling.

Having an easily-computable representation of the density distribution of the Galactic bar and its associated gravitational field is important for a number of applications, such as orbit integrations \citep{Stolte2014,Habing2016,Price-Whelan2016,Queiroz2020,Wylie2021}, hydrodynamical calculations of the response of the interstellar gas to a bar potential \citep{Armillotta2019,Tress2020,Sormani2020b}, and the study of the effect of the bar resonances \citep{Dehnen2000,Monari2019,Binney2020a,Chiba2021a}. However, there is a lack of a synthetic model that summarises in a concise way our current knowledge of the mass distribution of the Milky Way bar resulting from the huge body of work mentioned above. 

\cite{Portail2017a} (hereafter P17) constructed dynamical models of the Milky Way bar by integrating an N-body system and slowly adjusting the masses of the particles until the time-averaged density field and other model observables converged to prescribed data, using a made-to-measure (M2M) method \citep{Syer1996,deLorenzi2007}. The P17 models are constrained to reproduce a variety of stellar density and kinematic data, and they build upon previous reconstructions of the 3D bar density from red clump giant star counts \citep{Wegg2013,Wegg2015}. P17's overall best-fitting model had a pattern speed of $\Omega_{\rm p} = 40\kms\kpc^{-1}$. The pattern speed is one of the most important parameters of the bar since it sets the location of the resonances. More recently, there has been evidence for somewhat lower values of $\Omega_{\rm p}$ \citep{Clarke2019,Binney2020a,Chiba2021,Clarke2021}.  Here we consider the P17 model with $\Omega_{\rm p} = 37.5\kms\kpc^{-1}$ which is a good match to the VIRAC proper motions \citep{Clarke2019} and, with gas dynamical modelling, to the observed distribution of cold gas in the $(l,v)$-diagram \citep{Li2022}. This model can therefore be considered a state-of-the-art model that takes into account most of the available constraints on the structure of the MW bar. 

In this short research note we present a detailed 3D analytic fit to the stellar mass distribution of the dynamical model of P17. The analytic model can reproduce the 3D N-body density accurately, including the X-shape.
The analytic model and its associated gravitational potential are available as part of the software package {\sc Agama} for galactic dynamics \citep{Vasiliev2019}.

\section{The model} \label{sec:model}

The analytic model is composed by four components: three barred components, and an axisymmetric disc. The total density is:
\begin{equation} \label{eq:1}
\rho(x,y,z) = \underbrace{\rhobara + \rhobarb}_{\text{bar}} + \underbrace{\rhobarc}_{\text{long bar}}  + \rhodisc \,.
\end{equation}
The first two components together represent the bulge/bar (hereafter simply ``bar''), i.e. the X-shaped boxy component in the centre. The third component represents the long bar, i.e. a vertically flat extension of the bar which contribute to the ``ears", or bright enhancements at the ends, also known as ``ansae'' in the literature of external galaxies \cite[e.g.][]{Buta2013}. Note however that this decomposition is to some degree arbitrary and so these components only approximately correspond to the components with the same name in P17 and \cite{Wegg2015} (see also Sect.~\ref{sec:conclusion}).

The first term on the right-hand side of Equation~\eqref{eq:1} is a modification of equation 9 of \cite{Coleman2020}, which in turn is a generalisation of equation 10 of \cite{Freudenreich1998}:
\begin{equation}
\rhobara(x,y,z) = \rho_{1} \sech\left( a^{m} \right) \left[ 1+ \alpha \left(e^{-{a_+}^n}+e^{-{a_-}^n}\right) \right] e^{-\left( \frac{r}{r_{\rm cut}} \right)^2} \,,
\end{equation}
where
\begin{align}
a 		& = \left\{ \left[  \left(\frac{|x|}{x_1}\right)^{c_{\perp}} + \left(\frac{|y|}{y_1}\right)^{c_{\perp}} \right]^{\frac{c_\parallel}{c_\perp}} +  \left(\frac{|z|}{z_1}\right)^{c_{\parallel}} \right\}^{\frac{1}{c_\parallel}}\,, \\
a_{\pm}    & =  \left[ \left(\frac{x \pm c z}{x_c}\right)^2  + \left(\frac{y}{y_c}\right)^2  \right]^{\frac{1}{2}} \,, \\
r                & = \left( x^2 + y^2 + z^2 \right)^{\frac{1}{2}} \,.
\end{align}
The parameter $\alpha$ quantifies the strength of the X-shape, while the parameter $c$ quantifies its slope in the $(x,z)$ plane. 

The second and third terms on the right-hand side of Equation~\eqref{eq:1} have the same functional form, which is a modification of equation 9 of \cite{Wegg2015}:
\begin{align}
\rho_{\rm bar, i}(x,y,z) & = \rho_i \, e^{-a_i^{n_i}}  \sech^2 \left( \frac{z}{z_i} \right) e^{ - \left(\frac{R}{R_{\rm i, out}} \right)^{n_{\rm i, out}}} e^{- \left(\frac{R_{\rm i,in} }{R} \right)^{n_{\rm i,in}} } \,,
\end{align}
where $i = \{2,3\}$ and
\begin{align}
a_{i} & = \left[  \left(\frac{|x|}{x_{i}}\right)^{c_{\perp,i}} + \left(\frac{|y|}{y_i}\right)^{c_{\perp,i}} \right]^{\frac{1}{c_{\perp,i}}} \,, \\
R      & =  \left( x^2 + y^2 \right)^{\frac{1}{2}}  \label{eq:R} \,.
\end{align}
The disc is an axisymmetric component that covers the region outside the bar. We take its density distribution to be:
\begin{equation}
\rhodisc(R,z) = \frac{\Sigma_0}{4 z_d} e^{-\left(\frac{R}{R_d}\right)^{n_d}}e^{- \frac{R_{\rm cut}}{R}} \sech \left(\frac{|z|}{z_d}\right)^{m_d}  \,.
\end{equation}
where $R$ is the cylindrical radius given by Equation \eqref{eq:R}. 

We fit the multi-component density distribution given by \eqref{eq:1} to the time-averaged stellar density of the P17 M2M N-body model. The model of P17 originally has $10^6$ stellar particles; its time-averaged density was computed on the fly during the fitting run on a 3D grid with spacing $\Delta x = \Delta y = 0.3 \kpc$ and $\Delta z = 0.1 \kpc$. In this way the effective particle number is increased $\sim 100 \times$, allowing for a much smoother density distribution. We then minimise the quantity $\Delta^2 = \sum_i (\rho_{\text{analytic},i} - \rho_{ \text{N-body},i})^2$ using a standard Nelder-Mead algorithm, where $\rho_{\text{analytic},i}$ is the density of the analytic model, $\rho_{ \text{N-body},i}$ is the time-averaged density of the P17 N-body model, and the sum is extended over all points $i$ of the 3D grid. 

\begin{figure}
	\includegraphics[width=\columnwidth]{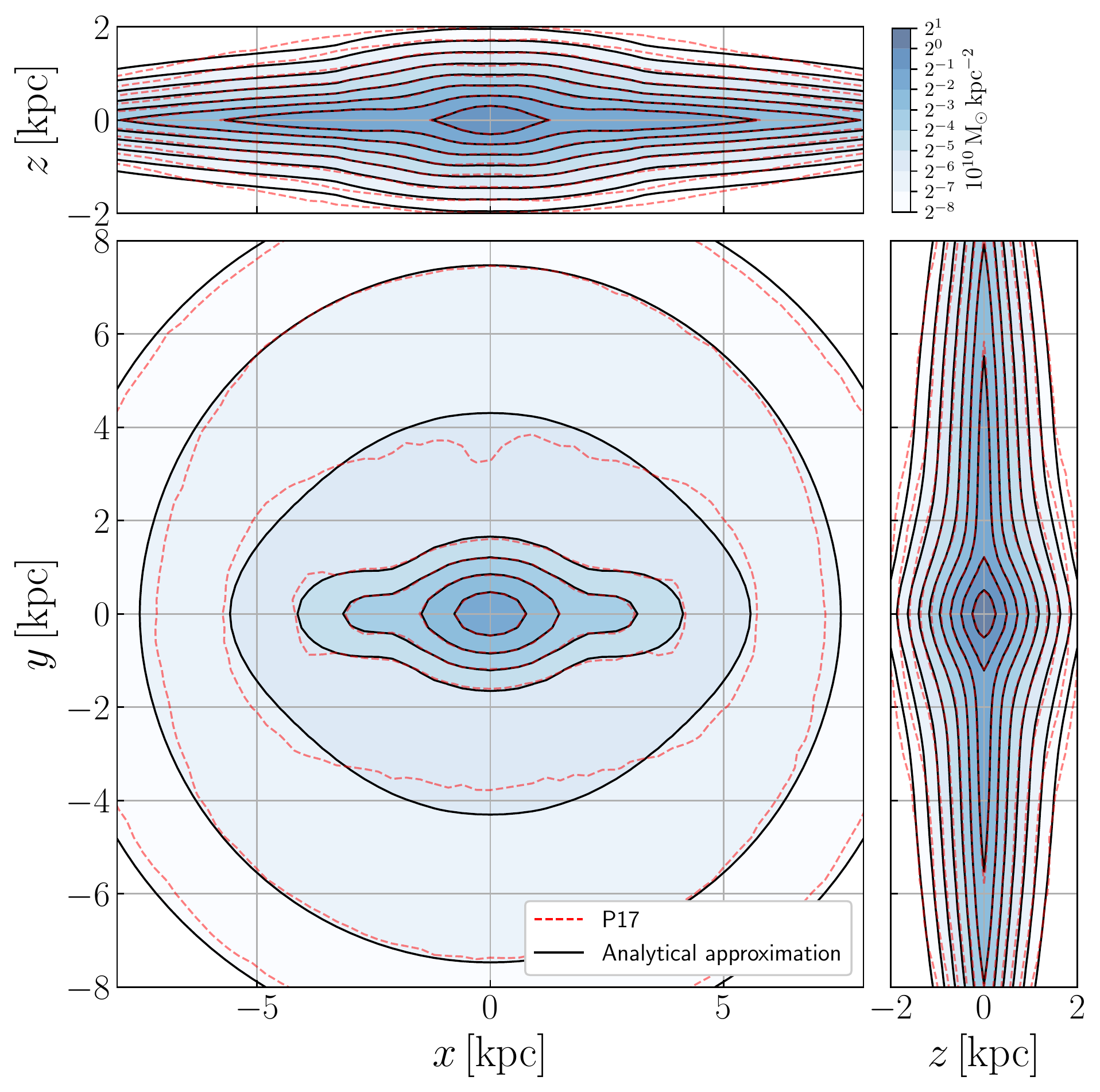}
    \caption{Surface density of the best-fitting analytic model compared to the time-averaged 3D density of the N-body model of P17.}
    \label{fig:1}
\end{figure}

\begin{figure}
\centering
\includegraphics[width=0.9\columnwidth]{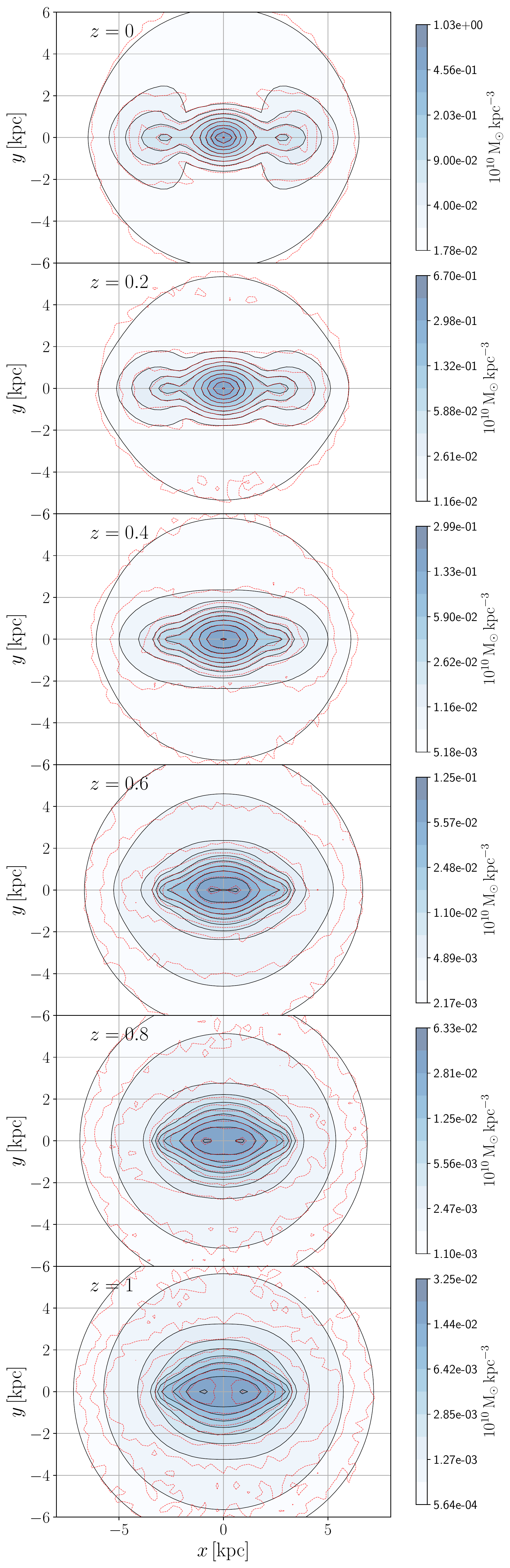}
    \caption{$(x,y)$ density slices at fixed values of $z$ of the analytic model (full black lines) compared to the P17 model (dashed red lines).}
    \label{fig:5}
\end{figure}

\begin{figure}
\centering
\includegraphics[width=0.8\columnwidth]{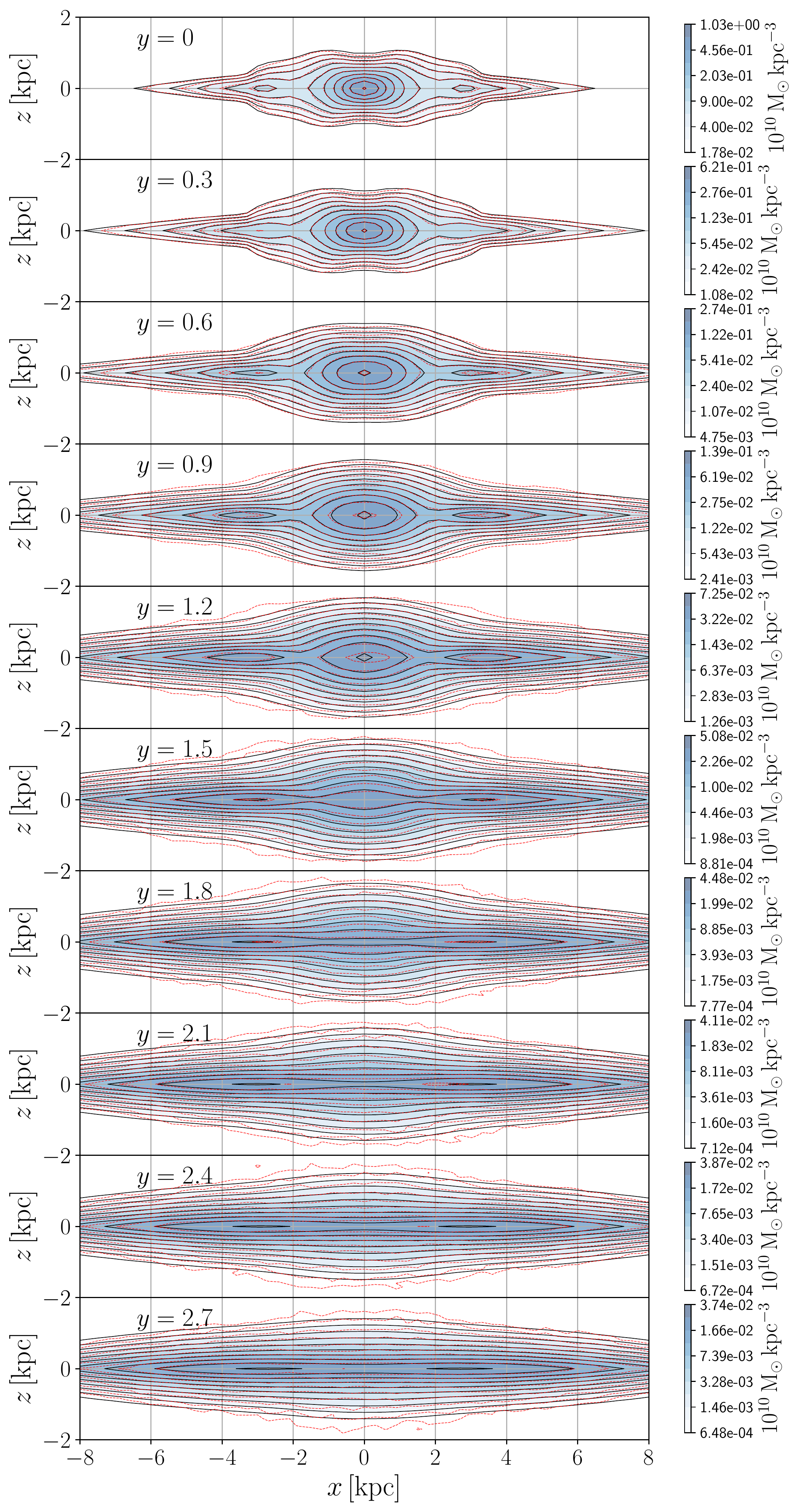}
    \caption{$(x,z)$ density slices at fixed values of $y$ of the analytic model (full black lines) compared to the P17 model (dashed red lines).}
    \label{fig:6}
\end{figure}

\setlength{\tabcolsep}{0.6ex}

\begin{table}
\begin{tabular}{lcr|lcr}  
\toprule
parameter 	& value	& 	units & parameter 	& value	& 	units \\
\midrule
     \multicolumn{3}{c}{\bf Barred component 1}  &     \multicolumn{3}{c}{\bf Disc}           \\
$\rho_1$    & 0.316        & $10^{10} \Msun \kpc^{-3} $ & $\Sigma_0$      & 0.103    & $10^{10} \Msun \kpc^{-2} $    \\
$x_1$         & 0.490        & $\kpc$                                       & $R_d$           & 4.754    & $\kpc$                        \\
$y_1$         & 0.392        & $\kpc$                                       & $z_d$           & 0.151    & $\kpc$                        \\
$z_1$         & 0.229        & $\kpc$                                       & $R_{\rm cut}$   & 4.688    & $\kpc$                        \\
$c_\parallel$ & 1.991        &                                               & $n_d$           & 1.536                                    \\
$c_\perp$     & 2.232        &                                              & $m_d$           & 0.716                                    \\
$m$           & 0.873        &                                              \\
$\alpha$      & 0.626        &                             \\
$n$           & 1.940        &                             \\
$c$           & 1.342        &                             \\
$x_c$         & 0.751        & $\kpc$                      \\
$y_c$         & 0.469        & $\kpc$                      \\
$r_{\rm cut}$ & 4.370        & $\kpc$                      \\
\midrule
     \multicolumn{3}{c}{\bf Barred component 2}  &   \multicolumn{3}{c}{\bf Barred component 3}           \\ 
$\rho_2$      &  0.050     & $10^{10} \Msun \kpc^{-3} $  & $\rho_3$      &  1743.049     & $10^{10} \Msun \kpc^{-3} $  \\
$x_2$           &  5.364     & $\kpc$                      & $x_3$           &  0.478     & $\kpc$                      \\
$y_2$           &  0.959     & $\kpc$                      & $y_3$           &  0.267     & $\kpc$                      \\
$z_2$           &  0.611     & $\kpc$                      & $z_3$           &  0.252     & $\kpc$                      \\
$n_2$           &  3.051     &                             & $n_3$           &  0.980                                   \\
$c_{\perp,2}$   &  0.970     &                             & $c_{\perp,3}$   &  1.879                                   \\
$R_{2,\rm out}$ &  3.190     & $\kpc$                      & $R_{3,\rm out}$ &  2.204     & $\kpc$                      \\
$R_{2, \rm in}$  &  0.558    & $\kpc$                      & $R_{3,\rm in}$  &  7.607     & $\kpc$                      \\
$n_{2,\rm out}$ &  16.731    &                               & $n_{3,\rm out}$ &  -27.291                                   \\
$n_{2,\rm in}$  &  3.196     &                             & $n_{3,\rm in}$  &  1.630                                   \\
\bottomrule
\end{tabular}
\caption{Parameters of the best-fitting analytic model.}
\label{tab:1}
\end{table}

\section{Results} \label{sec:results}

\begin{figure}
	\includegraphics[width=1.0\columnwidth]{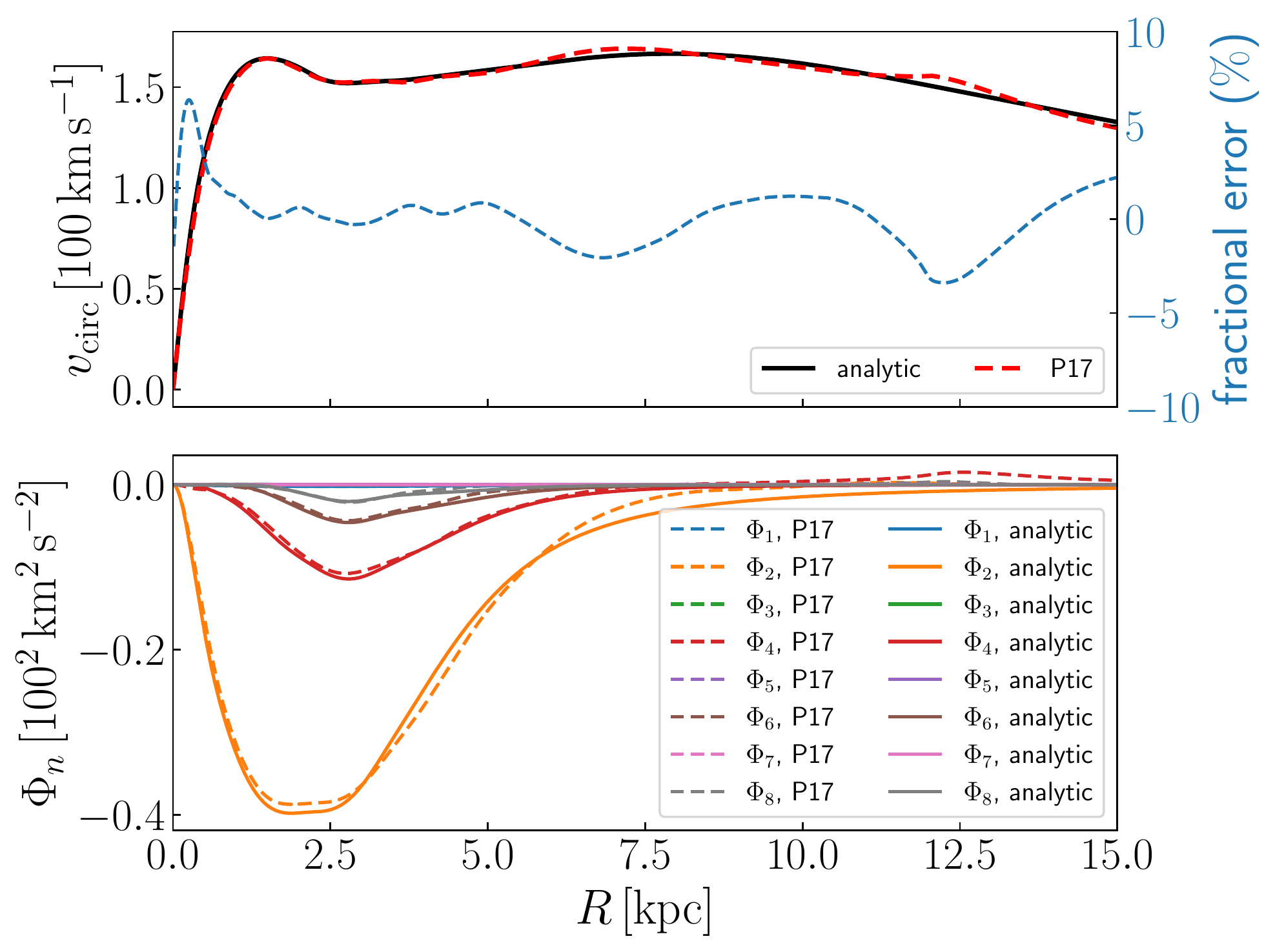}
    \caption{\emph{Top}: Circular velocity curve of the analytical model compared to P17. The blue dashed line shows the fractional error ($v_{\rm circ, analytic}-v_{\rm circ, P17})/v_{\rm circ, P17}$, which is of the order of few percent. \emph{Bottom}: multipoles of the potential in the $z=0$ plane, defined by expanding the potential as $\Phi(R,\theta) = \Phi_0(R) + \Phi_1(R)\cos(\theta) + \Phi_2(R)\cos(2\theta)+ \dots$, where $(R,\theta)$ are planar Polar coordinates. The circular velocity is defined as $v_{\rm circ}=\sqrt{R \di \Phi_0(R)/\di R}$.}
    \label{fig:7}
\end{figure}

\begin{figure*}
    \centering
	\includegraphics[width=0.9\textwidth]{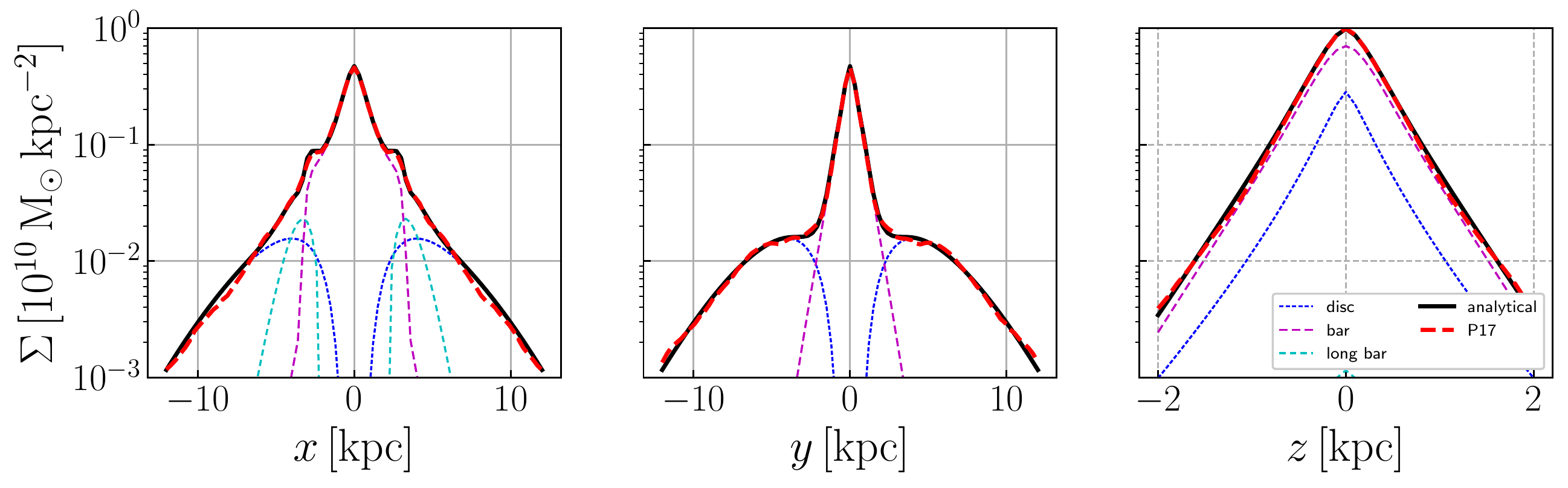}
    \caption{Surface density along the $x$, $y$ or $z$ axis of the analytic model (black full line) compared to the P17 made-to-measure N-body model (red dashed line). The left and middle panels correspond to the $x$ and $y$ axes in the $(x,y)$ map in Fig.~\ref{fig:1}, while the right panel correspond to the $z$ axis in the $(x,z)$ map.}
    \label{fig:2}
\end{figure*}

\begin{figure*}
 \centering
 \includegraphics[width=0.9\textwidth]{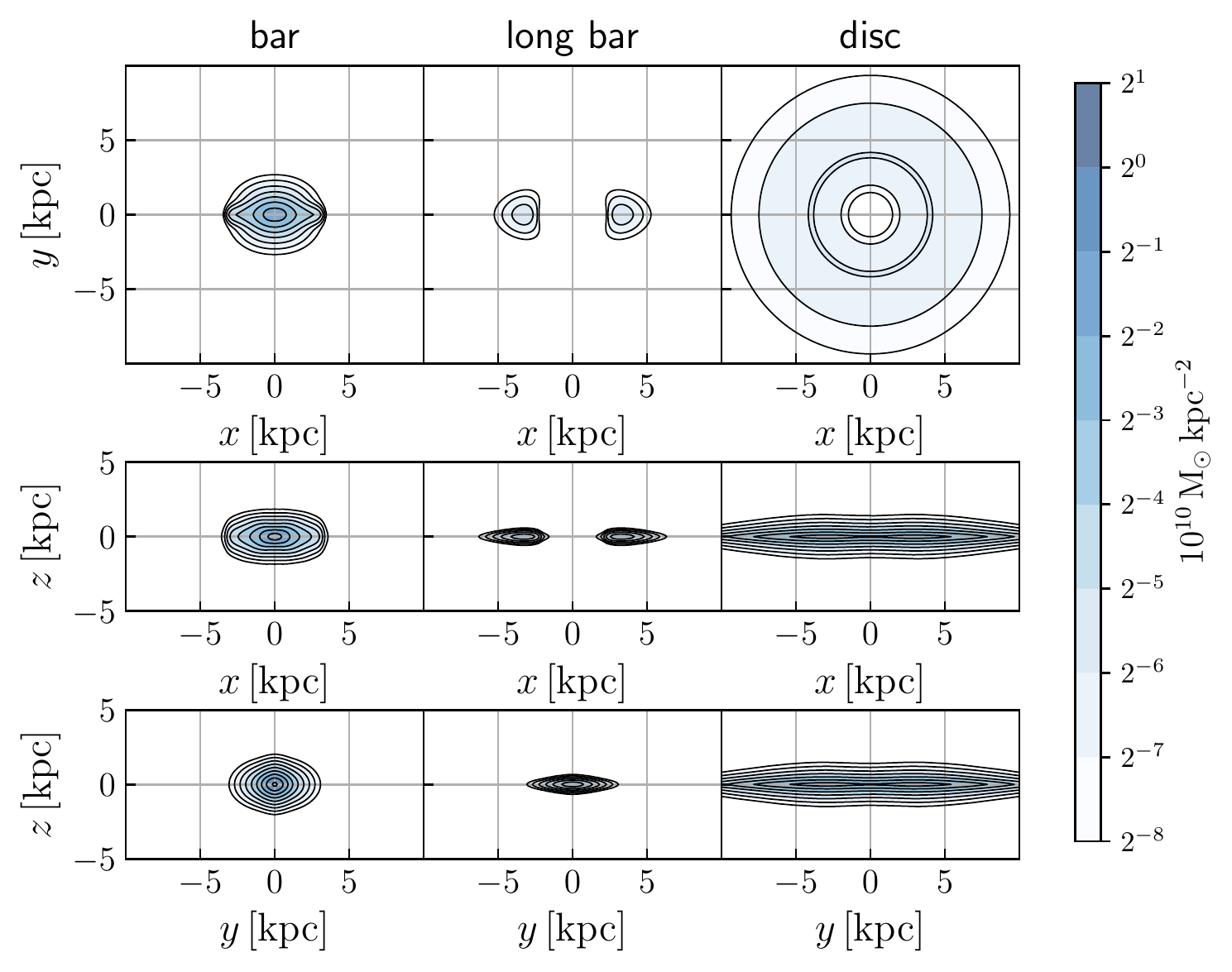}
 \caption{Surface density of the separate components of the analytical model.}
 \label{fig:4b}
\end{figure*}

The best-fitting parameters are reported in Table \ref{tab:1}. The statistical uncertainties on the parameters are negligibly small and not very meaningful, since those stemming from the chosen functional form of the density profiles are likely much larger. The total masses of the three barred components and of the disc are $M_{\text{bar,1}}=1.28 \times 10^{10} \Msun$, $M_{\text{bar,2}}=0.33 \times 10^{10} \Msun$, $M_{\text{bar,3}}=0.22 \times 10^{10} \Msun$ and $M_{\rm disc} = 3.19 \times 10^{10} \Msun$ respectively.  Figure \ref{fig:1} shows that the surface density of the analytic model provides an excellent fit to the surface density of the P17 N-body model. Figures~\ref{fig:5} and \ref{fig:6} shows that the analytic fit reproduces in detail the 3D structure of the N-body model, including the X-shape. Figure~\ref{fig:7} shows that the gravitational accelerations calculated with the analytic model reproduce well those of the P17 model, with an accuracy of the order of a few percent. The top panel shows the circular velocity curve, which quantifies the axisymmetric component of the bar and disc. The largest error ($\sim 5\%$) is observed at $R\sim 200 \pc$, where the potential is actually dominated by the nuclear stellar disc. The latter is not included in the present model and should be added to obtain the total gravitational field at $R \lesssim 300 \pc$ \citep{Sormani2022}. Note also that this circular velocity curve does not include the dark matter component. The bottom panel shows the first 8 multipoles, which quantify the non-axisymmetric part of the potential. These are also approximated well by the analytical model. We have also checked that the potential is very well approximated also outside the plane $z=0$. Finally, Fig.~\ref{fig:2} compares the surface density along the $x$, $y$ and $z$ axes, while Fig.~\ref{fig:4b} dissects the surface density of the analytical model into its separate components.

\section{Discussion and conclusion} \label{sec:conclusion}

We have presented an analytic model of the stellar mass distribution of the Milky Way bar, obtained by fitting a made-to-measure N-body model of P17. The analytic model reproduces the 3D density distribution of the N-body model in detail, including the X-shape. The model and the gravitational potential it generates are available as part of the software package Agama for galactic dynamics, and can be used for a number of applications, such as orbit integrations or hydrodynamical calculations of the response of the interstellar gas to a bar potential.

The distinction between ``bar'' and ``long bar'' in our fitting does not have an immediate physical meaning (e.g.\ in the sense that the two components correspond to two distinct orbital families, (\citealt{Skokos2002,Harsoula2009,Wylie2021}), and it should be merely seen as a convenient way of parametrising the density distribution. The bar + long bar components together can be considered as a meaningful component.

The P17 model was mainly fitted to data of the inner Galaxy and not to data of the outer disc. As such, the axisymmetric disc component might not represent the disc of the MW as accurately as other models available \citep[e.g.][]{McMillan2017}. Indeed, the disc of the P17 model produces a circular velocity that is slightly too low at the solar radius \citep{Li2022}. In some applications where a model of the gravitational potential of the Milky Way is needed, it might be convenient to replace the axisymmetric disc with a different model while keeping the bar + long bar components as presented here.

\section*{Acknowledgements}

MSC acknowledges support from the Deutsche Forschungsgemeinschaft (DFG) via the Collaborative Research Center (SFB 881, Project-ID 138713538) ``The Milky Way System'' (sub-projects A1, B1, B2 and B8) and from the European Research Council in the ERC Synergy Grant ``ECOGAL - Understanding our Galactic ecosystem: From the disk of the Milky Way to the formation sites of stars and planets'' (project ID 855130).

\section*{Data Availability}

The analytic fit and the associated gravitational potential are publicly available through the software package {\sc Agama} (\url{https://github.com/GalacticDynamics-Oxford/Agama}).


\bibliographystyle{mnras}
\bibliography{bibliography}



\bsp	
\label{lastpage}
\end{document}